\documentclass[12pt]{article}

\usepackage{epsfig,amsmath,amssymb,latexsym}

\newcommand{\be}{\begin{eqnarray}}
\newcommand{\ee}{\end{eqnarray}}
\newcommand{\bdm}{\begin{displaymath}}
\newcommand{\edm}{\end{displaymath}}
\newcommand{\ba}{\begin{array}}
\newcommand{\ea}{\end{array}}

\def\Orb{{\mathbf{S}^1/\mathbf{Z}_2}}
\def\mX{{\mathbf{X}}}

\numberwithin{equation}{section}

\begin{document}

\thispagestyle{empty}
\rightline{LMU-ASC 79/07}

\begin{center}

{\bf\Large Towards Dark Energy from String-Theory\footnote{Based on a talk given at DARK 2007, 6th International Heidelberg Conference on Dark Matter in Astro and Particle Physics, Sydney, Australia, 24-28 September 2007.}}

\vspace{1.3truecm}

Axel Krause\footnote{E-Mail: axel.krause@physik.uni-muenchen.de}

\vspace{.7truecm}

{\em Arnold Sommerfeld Center for Theoretical Physics\\
Department f\"ur Physik, Ludwig-Maximilians-Universit\"at M\"unchen\\
Theresienstr.~37, 80333 M\"unchen, Germany}

\end{center}

\vspace{1.0truecm}


\begin{abstract}
\noindent We discuss vacuum energy in string and M-theory with a focus on heterotic M-theory. In the latter theory a mechanism is described for maintaining zero vacuum energy {\em after} supersymmetry breaking. Higher-order corrections can be expected to give a sufficiently small amount of vacuum energy to possibly account for dark energy.
\end{abstract}

\newpage
\pagenumbering{arabic}

\section{The Dark Energy Enigma}

Cosmology underwent a revolution after it had been discovered in 1998 that our universe's current expansion accelerates, caused by some unknown, homogeneously distributed dark energy which dominates all other forms of energy or matter \cite{Riess:1998cb},
\cite{Perlmutter:1998np}. Under the assumption of a standard Friedmann-Robertson-Walker (FRW) cosmology, the dimming of distant type Ia supernovae, studies of angular anisotropies in the cosmic microwave background and studies of spatial correlations in the large-scale structure of galaxies have all led to this result. The price, on the other hand, for fitting the data without a dominating dark energy component is to adopt a spectrum of primordial density fluctuations which is not nearly scale-free, a Hubble constant which is globally lower than its locally measured value and a likely shift from the FRW to an inhomogeneous Lemaitre-Tolman-Bondi cosmology in order to explain the supernovae Ia Hubble diagram and the position of the baryon acoustic oscillation peak in the autocorrelation function of galaxies \cite{Sarkar:2007cx}. It seems therefore less problematic to accept the existence of dark energy in combination with the ordinary FRW description.

In contrast to this rather firm observational evidence for dark energy, there is little consensus on the theory side on what the correct explanation for the nature of dark energy could be. Over the past years various proposals for its origin have been made which include
\begin{itemize}
\item {\em cosmological constant} $\Lambda$:\\
leads to a time-independent energy-density $\rho_{DE}=\Lambda/8\pi G$, $\Lambda$ being the cosmological constant, and results in string-theory from metastable non-perturbative vacua \cite{Curio:2001qi}--\cite{Intriligator:2006dd}, warped compactifications with branes \cite{ArkaniHamed:2000eg}--\cite{Das:2007qn} or large extra dimensions \cite{Aghababaie:2003wz}--\cite{Burgess:2007ui}.

\item {\em quintessence}:\\
a dynamical scalar field rolls down an exponential potential and induces a time-dependent cosmological ``constant'' which evolves to small values at late times \cite{Wetterich:1987fm}--\cite{Wetterich:1994bg}. It opens up the possibility for tracker mechanisms \cite{Zlatev:1998tr}, which attempt to solve the coincidence problem, which arises from the near coincidence of the energy and the matter density in our present universe.

\item {\em holographic dark energy}:\\
to prevent a collapse into a black hole, the total energy of a region inside radius $L$ should not exceed the mass of a black hole with same radius $L$. This leads to a bound on the quantum zero-point energy density within a spherical region of radius $L$ \cite{Cohen:1998zx}. Saturation of this bound gives a relation between the UV cut-off and the IR cut-off of a quantum field theory which implies an energy density of the right magnitude \cite{Li:2004rb}--\cite{Setare:2006sv}. The actual equation of state depends, however, sensitively on the choice of the IR cut-off \cite{Hsu:2004ri}, \cite{Li:2004rb}.

\item {\em ghost cosmology}:\\
a ghost scalar $\phi$ has the wrong sign kinetic term and leads to an energy density unbounded from below. This renders the $\langle \phi \rangle = 0$ vacuum unstable. A non-minimal kinetic term could, however, allow the ghost to condense in a stable vacuum acquiring a non-zero constant velocity in field space \cite{ArkaniHamed:2003uy}. This new vacuum breaks Lorentz invariance spontaneously and leads to an infrared modification of gravity. Depending on the choice of the non-minimal kinetic term, the solutions to the coupled ghost-gravity field equations describe transitions from early power-law FRW scale-factors, $a(t) \sim t^{(2n-1)/3n}$, including radiation and matter cosmologies, to late time dark matter or dark energy dominated cosmologies \cite{Krause:2004bu}.

\item {\em modified gravity}:\\
an accelerated expansion of the universe can either have its origin in a cosmological constant resp.~scalar field added to the energy momentum tensor on the right hand side of the Einstein equations or in a modification of the geometrical part on its left hand side. The latter is the starting point for theories of modified gravity. For instance, one might add quadratic corrections \cite{Starobinsky:1980te} or inverse powers \cite{Capozziello:2003tk}--\cite{Carroll:2003wy} of the Ricci-scalar $R$ to the Einstein-Hilbert action or replace it by a general function $f(R)$ (see \cite{Nojiri:2006ri}, \cite{Woodard:2006nt} for reviews). Alternatively, brane-world constructions \cite{Akama:1982jy}--\cite{Randall:1999vf} can also modify the geometrical part of the Einstein equation. In the DGP model \cite{Dvali:2000hr} (for earlier work see \cite{Akama:1982jy}, \cite{Akama:1987ig}, \cite{Akama:2000vz}) quantum corrections induce a brane Einstein-Hilbert term next to a bulk Einstein-Hilbert term. This leads to an IR modification of gravity at lengths beyond a cross-over scale and gives rise to a late time acceleration.

\item {\em neutrino dark energy}:\\
the apparent closeness of the dark energy and the neutrino mass scale motivates a linkage between them. Concretely, one extends the Standard Model by singlet right-handed neutrinos and allows their Majorana masses to vary with the acceleron, a dynamical scalar field. The acceleron provides the link between a quintessential dark energy and the neutrino masses \cite{Hung:2000yg}--\cite{Fardon:2003eh}. See \cite{Bhatt:2007ah} for a recent update.
\end{itemize}

\section{Finetuning Problem and Two-Step Strategy}

In spite of the huge difference in the origins of the proposals for the nature of dark energy, they typically face one major problem. This is the required {\em tuning} or arbitrary choice of some parameters to generate the enormous smallness of the dark energy scale, $E_{DE} \simeq 1\text{meV}$, compared with any energy scale of fundamental physics such as the reduced Planck scale, $M_{Pl} = 2.4\times 10^{18}$GeV or the grand unification scale, $M_{GUT} = 2\times 10^{16}$GeV. Supersymmetry in particle theory or supergravity models alleviates the discrepancy but does not nullify it. The problem persists in having a vacuum energy scale, {\em after} supersymmetry breaking, which is of order the supersymmetry breaking scale, $M_{SUSY} \approx 1$TeV, and thus still far too large to match the dark energy scale. This is the ubiquitous cosmological constant (CC) problem. Since all forms of energy gravitate, it is not enough to tune a particular sector of the theory to give an $E_{DE}$ vacuum energy. Rather one has to ensure that at the same time none of the other sectors can develop energies surpassing $E_{DE}$. Moreover, in fundamental theories such as string-theory, low-energy parameters have their origin in dynamical scalar fields. Their values are thus determined dynamically and a fine-tuning is in principle unacceptable. Hence, the CC-problem cannot be glossed over in any proposal for dark energy and its dynamical solution is a necessary requisite for any successful explanation of dark energy.

The purpose of this article is to describe a {\em dynamical} way of generating the small dark energy scale. To this end, let us adopt
the following natural two-step approach to address the CC-problem
\begin{itemize}
\item First Step: find a mechanism which adjusts dynamically the vacuum energy {\em after} supersymmetry breaking to zero
\be
V_{vac} = 0 \; .
\ee
This should be true at leading orders in some suitable expansion, like a $1/M_{Pl}$ expansion in effective field theories or an $l_s = \sqrt{\alpha'}$ expansion in string-theory.

\item Second Step: higher-order perturbative and/or non-perturbative corrections to the theory are considered to lift the zero leading order vacuum energy to non-zero small positive values of the right magnitude
\be
V_{vac} \simeq E_{DE}^4 \; .
\ee
\end{itemize}
Of course, generating the right magnitude for $V_{vac}$ is not sufficient to explain dark energy. In addition, one has to check e.g.~that also the equation of state parameter $w$ lies sufficiently close to -1 to comply with observation. We won't discuss such additional checks further here, since already the generation of the right energy scale without tuning is a tremendous task and seems to be quite selective. The second step in this two-step approach is further motivated by various numerical coincidences which relate the observed dark energy scale to expressions which could plausibly arise from subleading perturbative or non-perturbative corrections. For instance, the well-known relation
\be
E_{DE}^4 \simeq M_{SUSY}^4 \times (M_{SUSY}/M_{Pl})^4 \; ,
\label{Vac1}
\ee
which gives the right size for $E_{DE}$ might arise at subleading order in a perturbative expansion in $M_{SUSY}/M_{Pl}$. Also, non-perturbative instanton corrections might produce
\be
E_{DE}^4 \simeq e^{-2/\alpha} M_{Pl}^4 \; ,
\label{Vac2}
\ee
where $\alpha\approx 1/137$ denotes the fine structure constant. A third numerical relation
\be
E_{DE}^4 \simeq e^{-E_{Pl}/2M_{GUT}} E_{Pl}^4 \; ,
\ee
where $E_{Pl}=1.2\times 10^{19}$GeV is the Planck energy, has been argued to arise in warped brane worlds \cite{Krause:2000gpa}, \cite{Krause:2000uj} and might also have a non-perturbative origin.
With this two-step approach in mind, let us in the sequel look at the prospects to realize it in supergravity, string-theory and finally M-theory.

\section{Dark Energy and Supergravity}

The best motivated theories beyond the Standard Model are based on supersymmetry, which we also adopt as one of the cornerstones in the following. Unbroken global supersymmetry has the attractive feature of enforcing a vanishing vacuum energy, as opposed to unbroken local supersymmetry. The latter being compatible with both zero and negative vacuum energies. The first question concerning supersymmetry is therefore whether it should be global or local. Here we note that a non-zero dark energy requires a curved spacetime, hence gravity and this implies local supersymmetry. Namely, in the presence of a non-zero homogeneous dark energy fluid, described by a diagonal energy-momentum tensor $T^{DE}_{\mu\nu}=\text{diag}(\rho,p,p,p)$, the Einstein field equations
\be
R_{\mu\nu} - \frac{1}{2}R g_{\mu\nu} = \frac{8\pi G}{c^4} T^{DE}_{\mu\nu}
\ee
have non-vanishing diagonal components on the right hand side which imply via the geometrical left hand side a curved spacetime and no flat Minkowski spacetime. It is thus mandatory that we work in local supersymmetry, i.e.~supergravity, which incorporates gravity.

If we then assume the phenomenologically favored $N=1$ supergravity framework in four dimensions, there are two options {\em before} breaking supersymmetry. Supersymmetry is either compatible with an anti de Sitter (AdS) or Minkowski spacetime. Both spacetimes are maximally symmetric. In view of the fact that the generated vacuum energy after supersymmetry breaking should not deviate much from zero, one would prefer to start off at this stage with a Minkowski solution. The AdS option would introduce already at this stage a large negative vacuum energy whose compensation requires an extremely precise fine-tuning. However, in the Minkowski case, it is the solution itself which cannot be obtained without fine-tuning in supergravity. The $N=1$ supersymmetry preserving vacua are characterized by a system of $N_c+1$ equations, which involve the superpotential $W$
\be
W = D_{\Phi_i} W = 0 \; , \qquad i=1,\hdots,N_c \; .
\ee
But there are only $N_c$ unknowns, the scalar components $\Phi_i$ of the chiral superfields. Since this system of equations is {\it overdetermined}, its solution is always {\it non-generic} and imposes fine-tuning on the physical parameters which enter the superpotential $W$ and/or the K\"ahler potential $K$ inside the K\"ahler covariant derivative $D_{\Phi_i} W = \partial_{\Phi_i} W + (\partial_{\Phi_i} K) W$. To deal with this fine-tuning there are two possibilities:
\begin{itemize}
\item one can accept the fine-tuning as unavoidable and a fundamental feature. In this case one might resort to a huge ``{\em landscape of vacua}'' combined with {\em anthropic} reasoning to try to make sense of a tiny, non-zero vacuum energy along the lines suggested in \cite{Weinberg:1987dv} (for a recent discussion see \cite{Bousso:2007gp}).

\item alternatively one could go {\em beyond the effective $N=1$ supergravity} in four dimensions and ask what string-theory has to offer in addition. After all supergravity captures only the massless spectrum of a string compactification and throws away all finite mass excitations.
\end{itemize}
Here, we decide to follow the latter route which brings us next to string-theory.

\section{Heterotic vs Type IIB String-Theory}

A priori there is a variety of five ten-dimensional string-theories which, together with eleven-dimensional supergravity, span the M-theory web. The five ten-dimensional perturbatively constructed string-theories represent special corners in this web at which the string couplings tends to zero. A generic point in this web, however, represents an eleven-dimensional theory, with the string coupling itself turning into the eleventh dimension when it grows.

To date, two regions have been identified in this M-theory web which are phenomenologically rich enough to admit a connection to the ``real'' cosmological and particle physics world. These are the ten-dimensional type IIB and the $E_8 \times E_8$ heterotic string theories. The latter is endowed with two $E_8$ gauge groups, one in a ``hidden'' sector and one in the ``visible'' sector. Both sectors interact only via (super)gravity. To bridge the gap from ten to four dimensions both string-theories have to be compactified on specific real 6-dimensional manifolds. For phenomenological reasons, and also for better technical control, one requires that this compactification preserves at least four supercharges in the effective four-dimensional supergravity theory. There is, however, an important distinction at this point. The type IIB theory has 32 supercharges in ten dimensions which is twice the amount of the $E_8\times E_8$ theory. This difference has the consequence that supersymmetric compactifications of the type IIB theory generically leave us with an AdS spacetime in four dimensions with a large negative energy density
\be
\text{Type IIB}: \qquad V_{vac} \ll 0 \; ,
\ee
whereas supersymmetric compactifications of the $E_8\times E_8$ heterotic theory lead to a Minkowski solution \cite{Strominger:1986uh} which has vanishing vacuum energy
\be
\text{Heterotic } E_8 \times E_8: \qquad V_{vac} = 0 \; .
\ee

As a consequence, to provide type IIB compactifications with a cosmologically relevant positive energy density, which can only be done by breaking supersymmetry, an additional ``uplift'' is required. This additional step, which in the simplest cases can be carried out by adding a supersymmetry-breaking anti-D-brane, adds positive energy density to the vacuum. In this approach one compensates the initially large negative AdS energy with the added positive ``uplift'' energy to end up with a desired but fine-tuned small vacuum energy. On the other hand, in the heterotic $E_8\times E_8$ compactifications, we start off at zero vacuum energy without any need for an ``uplift''. Furthermore, supersymmetry breaking can only generate a {\em positive} or zero vacuum energy thanks to the theory's perfect square potential \cite{Dine:1985rz}. The challenge in the heterotic theory is thus to break supersymmetry and keep the vacuum energy small. This will be our main concern in the remainder.

\section{Vacuum Energy after Supersymmetry Breaking in Heterotic String Compactifications}

With the just given motivation, let us investigate the vacuum energy after supersymmetry breaking in the heterotic string. The aim is to implement the first step of our initially advocated two-step approach, which is to break supersymmetry in such a way as to maintain at leading orders in $\alpha'$ the zero vacuum energy of the original supersymmetric theory. In fact such a mechanism had been proposed early on in the history of the heterotic string in \cite{Dine:1985rz}. The idea has been to consider a heterotic string compactification on a compact Calabi-Yau threefold $\mathbf{X}$ in the presence of 3-form Neveu-Schwarz flux $H$ and a gaugino condensate in the hidden sector $E_8$. This type of compactification yields a {\em positive definite} potential
\be
S_{pot} = -\frac{1}{2\kappa_{10}^2}
\int_{\phantom{}_{\mathbf{R^{1,3}}\times\mathbf{X}}}
\hspace{-8mm} e^{-\phi}
\Big(H-\frac{\alpha'}{16}e^{\phi/2}
\text{tr} \bar\chi \Gamma^{(3)} \chi \Big)^2 \; ,
\ee
with $\phi$ being the dilaton, $\kappa_{10}$ the ten-dimensional gravitational coupling constant, $\chi$ the ten-dimensional hidden sector gaugino and $\Gamma^{(3)}$ an antisymmetric three-index gamma matrix. The gaugino condensate $\langle \text{tr} \bar\chi \Gamma^{(3)} \chi\rangle$ can only assume values proportional to the Calabi-Yau's holomorphic three-form $\Omega$ and its complex conjugate
\be
\langle \text{tr} \bar\chi \Gamma^{(3)} \chi \rangle \sim \Lambda^3 \bar{\Omega} + \text{c.c.} \; ,
\ee
where $\Lambda^3$ represents the gaugino condensate in the effective four-dimensional theory, see below. This perfect square potential seems to relax dynamically towards zero vacuum energy by balancing the condensate with a non-zero $H$-flux. The resulting alignment of $\Omega$ and its complex conjugate with $H$
\be
H \sim \alpha' e^{\phi/2} (\Lambda^3 \bar{\Omega} + \text{c.c.})
\ee
{\em fixes all complex structure moduli} of the compactification. Furthermore, the 3-form flux $H$ must be of Hodge type $H^{(0,3)}$ and $H^{(3,0)}$ which {\em breaks supersymmetry}. We would thus be tempted to conclude that the first step has successfully been implemented in heterotic string compactifications, obtaining a vanishing vacuum energy up to order $(\alpha')^2$ despite breaking supersymmetry.

This is indeed a very attractive dynamical mechanism were it not for a quantum effect which poses a serious obstruction. Soon after the the above mechanism had been proposed in \cite{Dine:1985rz}, it was realized by Rohm and Witten that the 3-form flux $H$ had to be quantized to give a well-defined partition function \cite{Rohm:1985jv}. In the heterotic theory the exact form part $dB$ of the full 3-form flux, $H = dB + \frac{\alpha'}{4} (\Omega_L-\Omega_{YM})$, has to deliver quantized integer values when integrated over an arbitrary 3-cycle $\Sigma_3$
\be
\frac{1}{2\pi\alpha'} \int_{\Sigma_3} dB
= 2 \pi N \; , \quad N \in \mathbf{Z} \; .
\ee
Moreover, the integrated Yang-Mills and Lorentz Chern-Simons terms
\be
\frac{1}{8\pi} \int_{\Sigma_3}(\Omega_L-\Omega_{YM}) \; ,
\ee
which enter $H$ to render it gauge invariant, are only well defined modulo integers \cite{Rohm:1985jv}. Consequently, the {\em balancing equation}, which would result from minimizing the heterotic string perfect square potential, reads
\be
N \stackrel{?}{=} \frac{\alpha_0}{8\pi}
\big( \Lambda^3\bar\Pi + \bar\Lambda^3\Pi \big) \; ,
\label{HetStringBal}
\ee
after integration over some 3-cycle $\Sigma_3$. Here, $\alpha_0 = g^2_0/4\pi$ is the hidden sector gauge coupling, $\Pi = \int_{\Sigma_3} \Omega$ the period integral and
\be
\Lambda^3 = \langle \text{Tr} \lambda\lambda\rangle
= 16 \pi^2 M_{UV}^3 e^{-f_h/C_G}
\ee
the gaugino condensate of the hidden sector four-dimensional gaugino $\lambda$ with UV cut-off scale $M_{UV}$, hidden sector gauge kinetic function $f_h$ and dual Coxeter number $C_G$ related to the unbroken hidden sector gauge group $G$. The right hand side of the above balancing equation assumes exponentially small values due to the gaugino condensate. On the contrary, the left hand side assumes positive integers. It is therefore, in general, impossible to satisfy the heterotic string's balancing equation. In other words the quantization of $H$ prohibits the dynamical relaxation of the vacuum energy to zero.

This obstruction disappears when we go from the ten-dimensional heterotic string-theory to the eleven-dimensional heterotic M-theory, as we will discuss next, following \cite{Krause:2007gj}.

\section{Vacuum Energy after Supersymmetry Breaking in Heterotic M-Theory Compactifications}

\subsection{Flux Compactification Geometry}

\begin{figure}[bt]
\vspace{0.0cm}
\psfig{file=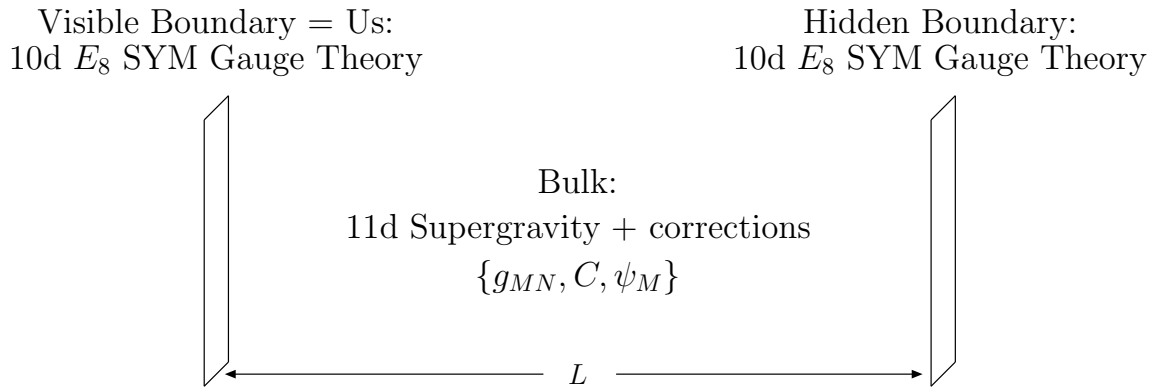,width=1.2\textwidth}
\vspace{-22cm}
\caption{\label{Setup} The eleven-dimensional heterotic M-theory setup.}
\vspace{1.5cm}
\end{figure}

The essential new ingredient concerning the vacuum energy problem, when going from the ten-dimensional heterotic string to the eleven-dimensional heterotic M-theory, is the warped geometrical background which extends along the extra eleventh dimension. Note that when the string coupling $g_s$ grows it turns into a geometrical entity, the size $L$ of the extra eleventh dimension $L \sim g_s^{2/3}$ (see fig.~\ref{Setup}).
The bulk eleven-dimensional spacetime in heterotic M-theory is bounded by two ten-dimensional boundaries, located at $x^{11}=0$ (visible boundary) and $x^{11}=L$ (hidden boundary), along the eleventh dimension $0\le x^{11}\le L$ which represents the orbifold $\Orb$. The visible boundary is assumed to host us together with the Standard Model and its grand unified extension at higher energies. The hidden boundary hosts a mirror world which is only capable to interact with us via (super)gravitational interactions which propagate through the bulk. A direct exploration of the hidden boundary through, say, electromagnetic waves is forbidden by the fact that the electromagnetic $U(1)$ gauge theory is part of the Standard Model gauge theory which resides exclusively on the visible boundary but has no extension into the eleven-dimensional bulk. Heterotic M-theory incorporates therefore the brane-world picture as an essential ingredient.

The physics in the bulk, at wavelengths larger than the 11-dimensional Planck-length, is governed by 11-dimensional supergravity. Its 4-form field-strength $G$ is sourced by the two $\Orb$ orbifold boundaries. They represent magnetic sources for $G$, much like a monopole represents a magnetic source for the ordinary electromagnetic gauge field-strength. The interesting consequence of this non-vanishing field-strength $G$ is that, via the Einstein equations, they cause the spacetime to be curved. More specifically, since we require a compactification which conserves an $N=1$ supersymmetry in the four large spacetime dimensions, it turns out that this spacetime has to be curved in a rather restricted way, namely by a warp-factor $e^{f(x^{11})}$ which depends only on the eleventh dimension $x^{11}$. Heterotic M-theory therefore operates on a spacetime which is described by a warped metric of the following type \cite{Curio:2000dw}--\cite{Curio:2003ur}
\be
ds^2 = e^{-2f(x^{11})} g_{\mu\nu} dx^\mu dx^\nu + e^{2f(x^{11})}
( g(\mathbf{X})_{lm} dy^l dy^m + (dx^{11})^2 ) \; .
\ee
The first part describes the 4-dimensional non-compact spacetime with metric $g_{\mu\nu}$, $\mu,\nu=0,\hdots,3$. The second and third part describe the Calabi-Yau metric $g(\mathbf{X})_{lm}$, $l,m=1,\hdots,6$ and the extension into the eleventh direction. The warp-factor is explicitly given by \cite{Curio:2000dw}--\cite{Curio:2003ur}
\be
e^{f(x^{11})} = |1-x^{11}Q_v|^{1/3} \; ,
\ee
where
\be
Q_v = -\frac{1}{8\pi V}
\Big( \frac{\kappa_{11}}{4\pi} \Big)^{2/3}
\int_{\mathbf{X}} J \wedge \big(\text{tr} F\wedge
F-\frac{1}{2}\text{tr} R\wedge R\big)
\ee
represents the charge with which the visible boundary couples to the supergravity 3-form potential $C$, $V$ denotes the unwarped Calabi-Yau volume, $J$ the K\"ahler-form of the Calabi-Yau $\mathbf{X}$ and $F$ resp.~$R$ the Yang-Mills resp.~Riemannian curvature 2-forms on the visible boundary. From this geometry it is easy to see that the volume of $\mathbf{X}$ decreases along the eleventh direction and vanishes eventually at a critical length
\be
L_c = 1/Q_v \; ,
\ee
given simply by the inverse of the charge. The warp-factor on the hidden boundary, i.e.~at $x^{11}=L$, which will play an important r\^ole later, can thus be expressed in geometrical terms as
\be
e^{f(L)} = \big(|L_c-L|/L_c\big)^{1/3} \; .
\ee
It clearly vanishes when the size of the eleventh dimension $L\le L_c$ reaches the critical length $L=L_c$.

With this warped background the question arises how does the warp-factor modify the heterotic perfect square potential? To answer this question, we must first clarify what becomes of the heterotic string perfect square potential when we go to heterotic M-theory. What happens is that the 3-form field-strength $H$ gets lifted to the 4-form field-strength $G$, and the two $E_8$ super Yang-Mills gauge sectors of the $E_8\times E_8$ gauge group become geometrically separated on the two boundaries (see fig.~\ref{Setup}). The warping causes the Calabi-Yau volume to decrease from visible to hidden boundary. Since the gauge couplings on the boundaries are inversely proportional to the Calabi-Yau volumes on those sites, it follows that the gauge theory on the hidden boundary is naturally strongly coupled when the visible boundary gauge theory is weakly coupled. Gaugino condensation will therefore occur naturally at high energies on the hidden boundary, and the perfect square potential turns into \cite{Horava:1996vs} (for technical reasons one is working on the $\mathbf{S}^1$ covering space of the $\Orb$ orbifold)
\be
S_{pot} = -\frac{1}{2\kappa^2_{11}}
\int_{\phantom{}_{\mathbf{R^{1,3}}\times\mathbf{X}\times\mathbf{S}^1}}
\hspace{-11mm}
\Big(G-\frac{\sqrt{2}}{32\pi} \left(\frac{\kappa_{11}}{4\pi}\right)^{2/3}
\text{tr} \bar\chi \hat\Gamma^{(3)} \chi \wedge \delta_L \Big)^2
\; .
\ee
The hat on $\hat\Gamma^{(3)}$ indicates that the 3-index gamma-matrix contains the vielbein of the warped metric, whereas $\delta_L$ denotes a Dirac-delta 1-form which localizes the gaugino condensate on the hidden boundary.

\subsection{Supersymmetry Breaking and Zero Vacuum Energy}

As in the heterotic string compactifications, we remain also with a perfect square potential in heterotic M-theory compactifications, which once more suggests that the system will relax dynamically towards zero potential energy. To see whether we can avoid the quantum obstruction which we faced for heterotic string compactifications, we have to analyze the balancing equation between $G$ and the condensate. In contrast to the heterotic string, we have now a warped background entering the perfect square in a non-trivial way. Let us first analyze the condensate. Here one finds, after a careful analysis carried out in \cite{Krause:2007gj}, that the condensate which is located on the hidden boundary at $x^{11}=L$, acquires the following warp-factor dependence
\be
\langle \text{tr} \bar\chi \hat\Gamma^{(3)} \chi \rangle
= 4\pi e^{-3f(L)} \alpha_0 \big( \Lambda^3
\bar\Omega + \bar\Lambda^3\Omega \big) \; .
\ee

Next, we have to take into account the quantization of the $G$ flux. In M-theory this is given by the following quantization condition \cite{Witten:1996md}
\be
\frac{1}{\sqrt{2}} \left(\frac{4\pi}{\kappa_{11}}\right)^{2/3}
\int_{\Sigma_4} G
+ \frac{\pi}{4} \int_{\Sigma_4} p_1(\mX)
= 2\pi N \; , \qquad N \in \mathbf{Z} \; .
\ee
The integration is performed over some arbitrary 4-cycle $\Sigma_4 \in H_4(\mX,\mathbf{Z})$ and the second term includes the first Pontryagin class $p_1(\mX)$ of $\mX$. The anticipated balancing of the $G$-flux with the condensate inside the perfect square tells us that $G$ must have one index along the eleventh dimension and being proportional to $\Omega$ resp.~its complex conjugate. The appropriate cycle over which we have to integrate to obtain a non-trivial result thus has to factorize like $\Sigma_4 = \Sigma_3 \times \mathbf{S}^1$, where $\Sigma_3$ is proportional to the Poincar\'e dual of $\Omega$ or $\bar\Omega$. Since this 4-cycle connects both boundaries, in principle, a further boundary contribution needs to be added to the above quantization condition \cite{Lukas:1997rb}. However, with $\Sigma_3$ being proportional to the Poincar\'e dual of $\Omega$ or $\bar\Omega$, both this boundary contribution and the $p_1(\mX)$ term in the above $G$-flux quantization condition do not contribute \cite{Lukas:1997rb}. We are thus left with the simple quantization rule
\be
\frac{1}{\sqrt{2}} \left(\frac{4\pi}{\kappa_{11}}\right)^{2/3}
\int_{\Sigma_3 \times \mathbf{S}^1} G
= 2\pi N \; , \qquad N \in \mathbf{Z} \; .
\ee

When we now apply this M-theory $G$-flux quantization to the balancing condition between $G$-flux and condensate at the minimum of the perfect square and integrate over the 4-cycle $\Sigma_3 \times \mathbf{S}^1$, we obtain the heterotic M-theory {\em balancing equation}
\be
e^{3f(L)} N
= \frac{\alpha_0}{8\pi}
\big( \Lambda^3\bar\Pi + \bar\Lambda^3\Pi \big) \; ,
\label{HetMBal}
\ee
where
\be
0\le e^{3f(L)} \le 1 \; .
\ee
Compared with the heterotic string balancing
equation~(\ref{HetStringBal}), the heterotic M-theory warped flux geometry led to the multiplication of the flux integer $N$ by the cube of the {\em warp-factor}, being evaluated at the position of the hidden boundary. This has the following important consequences
\begin{itemize}
\item the quantum obstruction to the balancing of the condensate with the flux ceases to exist. The {\em continuous} warp-factor suppresses the quantized flux. Moreover, the size $L$ of the $\Orb$ orbifold adjusts itself dynamically such as to set the perfect square potential to zero and satisfy eq.~(\ref{HetMBal}).

\item this dynamical balancing implies a {\em stabilization of $L$} close to $L_c$, leading to $e^{3f(L)} \ll 1$. This is in nice agreement with phenomenological constraints which favor a stabilization around $L_c$.

\item {\em supersymmetry is broken} since, as in the heterotic string case, $G$ becomes of Hodge type $H^{(0,3)}$ and $H^{(3,0)}$ along $\mathbf{X}$.
\end{itemize}
To summarize, we find {\em zero vacuum energy} after the system has evolved to the zero of its potential energy and supersymmetry has been broken. This accomplishes the first step of our two-step approach proposed in the beginning. Let us now comment on the second step.

\section{Dark Energy from Higher Order Corrections}

The perfect square potential in heterotic M-theory compactifications included contributions at order $\kappa_{11}^{2/3} \sim 1/M_{11}^3$ and $(\kappa_{11}^{2/3})^2 \sim 1/M_{11}^6$, where $M_{11}$ is the eleven-dimensional Planck mass. There is, however, no theorem which would protect the perfect square structure up to higher orders $(\kappa_{11}^{2/3})^n \sim 1/M_{11}^{3n}$, $n\ge 3$. These higher order contributions are suppressed by corresponding powers of the four-dimensional Planck mass in the effective four-dimensional theory and could thus generate a sufficiently small residual vacuum energy. One might therefore hope that perhaps a relation like eq.~(\ref{Vac1}) might become derivable.

One should also add the known $R^4$ and their superpartner $C\wedge R^4$ corrections to the heterotic M-theory action. This leads to small corrections to the heterotic M-theory flux compactification geometry \cite{Anguelova:2005jr}. It is clear that this corrected background solution, being a supersymmetric solution to the field equations, must give itself a vanishing vacuum energy in the absence of supersymmetry breaking, as has been demonstrated in detail for the leading order warped background in \cite{Curio:2003ur}. So,
it's again the task of the gaugino condensate and the induced $G$-flux to break supersymmetry and generate a non-zero vacuum energy. Since both the condensate and the $G$-flux reside on the hidden boundary, where the warp-factor is smallest, higher order corrections to the perfect square potential which contribute to the vacuum energy can be expected to come with strong suppressions by high powers of the small warp-factor $e^{f(L)}$. Moreover, these higher order corrections are suppressed through additional powers of $1/M_{11}$. There has been recent progress in obtaining a heterotic M-theory action to all orders in $\kappa_{11}^{2/3}$
\cite{Moss:2003bk}--\cite{Moss:2005zw}. This action would allow a quantitative study of the vacuum energy incorporating these higher order corrections to the vacuum energy and is presently under investigation.

\section{Summary}

We can therefore conclude that
\begin{itemize}
\item dynamical supersymmetry breaking is possible in heterotic M-theory compactifications with zero vacuum energy at leading orders in $\kappa_{11}^{2/3}$

\item higher-order corrections can be expected to give a small correction to the vanishing leading order vacuum energy and could thus account for the observed dark energy

\item the quantum obstruction of the heterotic string to obtain zero vacuum energy after supersymmetry breaking is avoided in heterotic M-theory through a dependence on the continuous warp-factor which is not available in the heterotic string

\item the $\Orb$ size $L$, i.e.~the dilaton in M-theory disguise, is stabilized close to the critical length $L_c$, an important fact for phenomenological and cosmological applications
\cite{Witten:1996mz}--\cite{Curio:2006dc}

\item complex structure moduli are stabilized by $G$-flux which is generated as a dynamic response to the hidden sector gaugino condensate at the minimum of the perfect square potential
\end{itemize}
Even though it was not our focus here, let us finally comment on dark matter in the same framework. Moduli stabilization of heterotic M-theory compactifications favor a broken hidden $E_8$ gauge group \cite{Becker:2004gw}. This implies hidden matter which can interact with the visible matter only via bulk (super)gravitational interactions and has therefore been considered as dark matter candidates \cite{Krause:2004up}. A complete analysis of the viability of this class of dark matter candidates still needs to be done but, if successful, this type of dark matter would nicely complement the outlined picture for the possible M-theory origin of dark energy.

\newpage
\bigskip
\noindent {\large \bf Acknowledgements}

\noindent We thank the organizers of DARK 2007 for a wonderful conference and acknowledge support from the DFG and the Transregional Collaborative Research Centre TRR~33 ``The Dark Universe''.


\begin{thebibliography}{99}

\bibitem{Riess:1998cb}
  A.~G.~Riess {\it et al.}  [Supernova Search Team Collaboration],
  Astron.\ J.\  {\bf 116}, 1009 (1998)
  [arXiv:astro-ph/9805201].

\bibitem{Perlmutter:1998np}
  S.~Perlmutter {\it et al.}  [Supernova Cosmology Project Collaboration],
  Astrophys.\ J.\  {\bf 517}, 565 (1999)
  [arXiv:astro-ph/9812133].

\bibitem{Sarkar:2007cx}
  S.~Sarkar,
  arXiv:0710.5307 [astro-ph].

\bibitem{Curio:2001qi}
  G.~Curio and A.~Krause,
  Nucl.\ Phys.\  B {\bf 643}, 131 (2002)
  [arXiv:hep-th/0108220].

\bibitem{Kachru:2003aw}
  S.~Kachru, R.~Kallosh, A.~Linde and S.~P.~Trivedi,
  Phys.\ Rev.\  D {\bf 68}, 046005 (2003)
  [arXiv:hep-th/0301240].

\bibitem{Intriligator:2006dd}
  K.~Intriligator, N.~Seiberg and D.~Shih,
  JHEP {\bf 0604}, 021 (2006)
  [arXiv:hep-th/0602239].

\bibitem{ArkaniHamed:2000eg}
  N.~Arkani-Hamed, S.~Dimopoulos, N.~Kaloper and R.~Sundrum,
  Phys.\ Lett.\  B {\bf 480}, 193 (2000)
  [arXiv:hep-th/0001197].

\bibitem{Kachru:2000hf}
  S.~Kachru, M.~B.~Schulz and E.~Silverstein,
  Phys.\ Rev.\  D {\bf 62}, 045021 (2000)
  [arXiv:hep-th/0001206].

\bibitem{Tye:2000fw}
  S.~H.~H.~Tye and I.~Wasserman,
  Phys.\ Rev.\ Lett.\  {\bf 86}, 1682 (2001)
  [arXiv:hep-th/0006068].

\bibitem{Krause:2000gpa}
  A.~Krause,
  Nucl.\ Phys.\  B {\bf 748}, 98 (2006)
  [arXiv:hep-th/0006226].

\bibitem{Krause:2000uj}
  A.~Krause,
  JHEP {\bf 0309}, 016 (2003)
  [arXiv:hep-th/0007233].

\bibitem{Kehagias:2000dg}
  A.~Kehagias and K.~Tamvakis,
  Mod.\ Phys.\ Lett.\  A {\bf 17}, 1767 (2002)
  [arXiv:hep-th/0011006].

\bibitem{Cline:2000ky}
  J.~M.~Cline and H.~Firouzjahi,
  Phys.\ Lett.\  B {\bf 514}, 205 (2001)
  [arXiv:hep-ph/0012090].

\bibitem{Tetradis:2000fu}
  N.~Tetradis,
  Phys.\ Lett.\  B {\bf 509}, 307 (2001)
  [arXiv:hep-th/0012106].

\bibitem{Nilles:2003km}
  H.~P.~Nilles, A.~Papazoglou and G.~Tasinato,
  Nucl.\ Phys.\  B {\bf 677}, 405 (2004)
  [arXiv:hep-th/0309042].

\bibitem{Kamani:2006tv}
  D.~Kamani,
  arXiv:hep-th/0611339.

\bibitem{Park:2007ij}
  E.~K.~Park and P.~S.~Kwon,
  JHEP {\bf 0711}, 051 (2007)
  [arXiv:hep-th/0702171].

\bibitem{Das:2007qn}
  S.~Das, D.~Maity and S.~SenGupta,
  arXiv:0711.1744 [hep-th].

\bibitem{Aghababaie:2003wz}
  Y.~Aghababaie, C.~P.~Burgess, S.~L.~Parameswaran and F.~Quevedo,
  Nucl.\ Phys.\  B {\bf 680}, 389 (2004)
  [arXiv:hep-th/0304256].

\bibitem{Burgess:2004yq}
  C.~P.~Burgess, J.~Matias and F.~Quevedo,
  Nucl.\ Phys.\  B {\bf 706}, 71 (2005)
  [arXiv:hep-ph/0404135].

\bibitem{Burgess:2007ui}
  C.~P.~Burgess,
  arXiv:0708.0911 [hep-ph].

\bibitem{Wetterich:1987fm}
  C.~Wetterich,
  Nucl.\ Phys.\  B {\bf 302}, 668 (1988).

\bibitem{Ratra:1987rm}
  B.~Ratra and P.~J.~E.~Peebles,
  Phys.\ Rev.\  D {\bf 37}, 3406 (1988).

\bibitem{Wetterich:1994bg}
  C.~Wetterich,
  Astron.\ Astrophys.\  {\bf 301}, 321 (1995)
  [arXiv:hep-th/9408025].

\bibitem{Zlatev:1998tr}
  I.~Zlatev, L.~M.~Wang and P.~J.~Steinhardt,
  Phys.\ Rev.\ Lett.\  {\bf 82}, 896 (1999)
  [arXiv:astro-ph/9807002].

\bibitem{Cohen:1998zx}
  A.~G.~Cohen, D.~B.~Kaplan and A.~E.~Nelson,
  Phys.\ Rev.\ Lett.\  {\bf 82}, 4971 (1999)
  [arXiv:hep-th/9803132].

\bibitem{Li:2004rb}
  M.~Li,
  Phys.\ Lett.\  B {\bf 603}, 1 (2004)
  [arXiv:hep-th/0403127].

\bibitem{Gong:2004fq}
  Y.~g.~Gong,
  Phys.\ Rev.\  D {\bf 70}, 064029 (2004)
  [arXiv:hep-th/0404030].

\bibitem{Horvat:2004vn}
  R.~Horvat,
  Phys.\ Rev.\  D {\bf 70}, 087301 (2004)
  [arXiv:astro-ph/0404204].

\bibitem{Huang:2004ai}
  Q.~G.~Huang and M.~Li,
  JCAP {\bf 0408}, 013 (2004)
  [arXiv:astro-ph/0404229].

\bibitem{Enqvist:2004xv}
  K.~Enqvist and M.~S.~Sloth,
  Phys.\ Rev.\ Lett.\  {\bf 93}, 221302 (2004)
  [arXiv:hep-th/0406019].

\bibitem{Gong:2004cb}
  Y.~g.~Gong, B.~Wang and Y.~Z.~Zhang,
  Phys.\ Rev.\  D {\bf 72}, 043510 (2005)
  [arXiv:hep-th/0412218].

\bibitem{Myung:2004ch}
  Y.~S.~Myung,
  Phys.\ Lett.\  B {\bf 610}, 18 (2005)
  [arXiv:hep-th/0412224].

\bibitem{Guberina:2005mp}
  B.~Guberina, R.~Horvat and H.~Nikolic,
  Phys.\ Rev.\  D {\bf 72}, 125011 (2005)
  [arXiv:astro-ph/0507666].

\bibitem{Setare:2006wh}
  M.~R.~Setare,
  Phys.\ Lett.\  B {\bf 642}, 1 (2006)
  [arXiv:hep-th/0609069].

\bibitem{Setare:2006sv}
  M.~R.~Setare,
  Phys.\ Lett.\  B {\bf 642}, 421 (2006)
  [arXiv:hep-th/0609104].

\bibitem{Hsu:2004ri}
  S.~D.~H.~Hsu,
  Phys.\ Lett.\  B {\bf 594}, 13 (2004)
  [arXiv:hep-th/0403052].

\bibitem{ArkaniHamed:2003uy}
  N.~Arkani-Hamed, H.~C.~Cheng, M.~A.~Luty and S.~Mukohyama,
  JHEP {\bf 0405}, 074 (2004)
  [arXiv:hep-th/0312099].

\bibitem{Krause:2004bu}
  A.~Krause and S.~P.~Ng,
  Int.\ J.\ Mod.\ Phys.\  A {\bf 21}, 1091 (2006)
  [arXiv:hep-th/0409241].

\bibitem{Starobinsky:1980te}
  A.~A.~Starobinsky,
  Phys.\ Lett.\  B {\bf 91}, 99 (1980).

\bibitem{Capozziello:2003tk}
  S.~Capozziello, S.~Carloni and A.~Troisi,
  arXiv:astro-ph/0303041.

\bibitem{Capozziello:2003gx}
  S.~Capozziello, V.~F.~Cardone, S.~Carloni and A.~Troisi,
  Int.\ J.\ Mod.\ Phys.\  D {\bf 12}, 1969 (2003)
  [arXiv:astro-ph/0307018].

\bibitem{Carroll:2003wy}
  S.~M.~Carroll, V.~Duvvuri, M.~Trodden and M.~S.~Turner,
  Phys.\ Rev.\  D {\bf 70}, 043528 (2004)
  [arXiv:astro-ph/0306438].

\bibitem{Nojiri:2006ri}
  S.~Nojiri and S.~D.~Odintsov,
  Int.\ J.\ Geom.\ Meth.\ Mod.\ Phys.\  {\bf 4}, 115 (2007)
  [arXiv:hep-th/0601213].

\bibitem{Woodard:2006nt}
  R.~P.~Woodard,
  Lect.\ Notes Phys.\  {\bf 720}, 403 (2007)
  [arXiv:astro-ph/0601672].

\bibitem{Akama:1982jy}
  K.~Akama,
  Lect.\ Notes Phys.\  {\bf 176}, 267 (1982)
  [arXiv:hep-th/0001113].

\bibitem{Rubakov:1983bb}
  V.~A.~Rubakov and M.~E.~Shaposhnikov,
  Phys.\ Lett.\  B {\bf 125}, 136 (1983).

\bibitem{Visser:1985qm}
  M.~Visser,
  Phys.\ Lett.\  B {\bf 159}, 22 (1985)
  [arXiv:hep-th/9910093].

\bibitem{Horava:1996ma}
  P.~Horava and E.~Witten,
  Nucl.\ Phys.\  B {\bf 475}, 94 (1996)
  [arXiv:hep-th/9603142].

\bibitem{ArkaniHamed:1998rs}
  N.~Arkani-Hamed, S.~Dimopoulos and G.~R.~Dvali,
  Phys.\ Lett.\  B {\bf 429}, 263 (1998)
  [arXiv:hep-ph/9803315].

\bibitem{Randall:1999vf}
  L.~Randall and R.~Sundrum,
  Phys.\ Rev.\ Lett.\  {\bf 83}, 4690 (1999)
  [arXiv:hep-th/9906064].

\bibitem{Dvali:2000hr}
  G.~R.~Dvali, G.~Gabadadze and M.~Porrati,
  Phys.\ Lett.\  B {\bf 485}, 208 (2000)
  [arXiv:hep-th/0005016].

\bibitem{Akama:1987ig}
  K.~Akama,
  Prog.\ Theor.\ Phys.\  {\bf 78}, 184 (1987).

\bibitem{Akama:2000vz}
  K.~Akama and T.~Hattori,
  Mod.\ Phys.\ Lett.\  A {\bf 15}, 2017 (2000)
  [arXiv:hep-th/0008133].

\bibitem{Hung:2000yg}
  P.~Q.~Hung,
  arXiv:hep-ph/0010126.

\bibitem{Gu:2003er}
  P.~Gu, X.~Wang and X.~Zhang,
  Phys.\ Rev.\  D {\bf 68}, 087301 (2003)
  [arXiv:hep-ph/0307148].

\bibitem{Fardon:2003eh}
  R.~Fardon, A.~E.~Nelson and N.~Weiner,
  JCAP {\bf 0410}, 005 (2004)
  [arXiv:astro-ph/0309800].

\bibitem{Bhatt:2007ah}
  J.~R.~Bhatt, P.~H.~Gu, U.~Sarkar and S.~K.~Singh,
  arXiv:0711.2728 [hep-ph].

\bibitem{Weinberg:1987dv}
  S.~Weinberg,
  Phys.\ Rev.\ Lett.\  {\bf 59}, 2607 (1987).

\bibitem{Bousso:2007gp}
  R.~Bousso,
  arXiv:0708.4231 [hep-th].

\bibitem{Strominger:1986uh}
  A.~Strominger,
  Nucl.\ Phys.\  B {\bf 274}, 253 (1986).

\bibitem{Dine:1985rz}
  M.~Dine, R.~Rohm, N.~Seiberg and E.~Witten,
  Phys.\ Lett.\  B {\bf 156}, 55 (1985).

\bibitem{Rohm:1985jv}
  R.~Rohm and E.~Witten,
  Annals Phys.\  {\bf 170}, 454 (1986).

\bibitem{Krause:2007gj}
  A.~Krause,
  Phys.\ Rev.\ Lett.\  {\bf 98}, 241601 (2007)
  [arXiv:hep-th/0701009].

\bibitem{Curio:2000dw}
  G.~Curio and A.~Krause,
  Nucl.\ Phys.\  B {\bf 602}, 172 (2001)
  [arXiv:hep-th/0012152].

\bibitem{Krause:2001qf}
  A.~Krause,
  Fortsch.\ Phys.\  {\bf 49}, 163 (2001).

\bibitem{Curio:2003ur}
  G.~Curio and A.~Krause,
  Nucl.\ Phys.\  B {\bf 693}, 195 (2004)
  [arXiv:hep-th/0308202].

\bibitem{Horava:1996vs}
  P.~Horava,
  Phys.\ Rev.\  D {\bf 54}, 7561 (1996)
  [arXiv:hep-th/9608019].

\bibitem{Witten:1996md}
  E.~Witten,
  J.\ Geom.\ Phys.\  {\bf 22}, 1 (1997)
  [arXiv:hep-th/9609122].

\bibitem{Lukas:1997rb}
  A.~Lukas, B.~A.~Ovrut and D.~Waldram,
  Phys.\ Rev.\ D {\bf 57}, 7529 (1998)
  [arXiv:hep-th/9711197].

\bibitem{Anguelova:2005jr}
  L.~Anguelova and D.~Vaman,
  Nucl.\ Phys.\  B {\bf 733}, 132 (2006)
  [arXiv:hep-th/0506191].

\bibitem{Moss:2003bk}
  I.~G.~Moss,
  Phys.\ Lett.\  B {\bf 577}, 71 (2003)
  [arXiv:hep-th/0308159].

\bibitem{Moss:2004ck}
  I.~G.~Moss,
  Nucl.\ Phys.\  B {\bf 729}, 179 (2005)
  [arXiv:hep-th/0403106].

\bibitem{Moss:2005zw}
  I.~G.~Moss,
  Phys.\ Lett.\  B {\bf 637}, 93 (2006)
  [arXiv:hep-th/0508227].

\bibitem{Witten:1996mz}
  E.~Witten,
  Nucl.\ Phys.\  B {\bf 471}, 135 (1996)
  [arXiv:hep-th/9602070].

\bibitem{Banks:1996ss}
  T.~Banks and M.~Dine,
  Nucl.\ Phys.\  B {\bf 479}, 173 (1996)
  [arXiv:hep-th/9605136].

\bibitem{Becker:2004gw}
  M.~Becker, G.~Curio and A.~Krause,
  Nucl.\ Phys.\  B {\bf 693}, 223 (2004)
  [arXiv:hep-th/0403027].

\bibitem{Becker:2005sg}
  K.~Becker, M.~Becker and A.~Krause,
  Nucl.\ Phys.\  B {\bf 715}, 349 (2005)
  [arXiv:hep-th/0501130].

\bibitem{Becker:2005pv}
  K.~Becker, M.~Becker and A.~Krause,
  Phys.\ Rev.\  D {\bf 74}, 045023 (2006)
  [arXiv:hep-th/0510066].

\bibitem{Curio:2006dc}
  G.~Curio and A.~Krause,
  Phys.\ Rev.\  D {\bf 75}, 126003 (2007)
  [arXiv:hep-th/0606243].

\bibitem{Krause:2004up}
  A.~Krause,
  in the proceedings of SUSY 2003, 11th Annual International Conference on Supersymmetry and the Unification of Fundamental Interactions, Tucson, Arizona, [arXiv:hep-th/0404001].

\end{thebibliography}
\end{document}